\documentclass[a4paper]{aa}
\usepackage{psfig,times}

\begin{document}

\newif\ifAMStwofonts
\def\ni{\noindent} 
\def\ea{{et\thinspace al.}\ }                       
\def\eg{{e.g.}\ }                                
\def\ie{{i.e.}\ }                               
\def\cf{{cf.}\ } 
\def\rot{\mathop{\rm rot}\nolimits}
\def\div{\mathop{\rm div}\nolimits} 
\renewcommand{\vec}[1]{\mbox{\boldmath $#1$}}
 
\def\solar{\ifmode_{\mathord\odot}\else$_{\mathord\odot}$\fi} 
\def\gsim{\lower.4ex\hbox{$\;\buildrel >\over{\scriptstyle\sim}\;$}} 
\def\lsim{\lower.4ex\hbox{$\;\buildrel <\over{\scriptstyle\sim}\;$}} 
\def\Ca{Ca\thinspace {\rm II}} 
\def\~  {$\sim$} 
\def\cl {\centerline} 
\def\rl {\rightline} 
\def\x	{\times} 
 
\def\bl{\par\vskip 12pt\noindent} 
\def\bll{\par\vskip 24pt\noindent} 
\def\blll{\par\vskip 36pt\noindent}
\def\alf{$\alpha$}
\def\L{$\Lambda$}
\def\Om{$\Omega$}
\def\nT{$\nu_{\rm T}$\ }
\def\mT{$\mu_{\rm T}$\ }
\def\cT{$\chi_{\rm T}$\ }

\def\apj{{ApJ}}       
\def\apjs{{ Ap. J. Suppl.}} 
\def\apjl{{ Ap. J. Letters}} 
\def\pasp{{ Pub. A.S.P.}} 
\def\mn{{MNRAS}} 
\def\aa{{A\&A}} 
\def\aasup{{ Astr. Ap. Suppl.}} 
\def\baas{{ Bull. A.A.S.}\ } 
\def\csss{{Cool Stars, Stellar Systems, and the Sun}}
\def\an{{Astron. Nachr.}}
\def\sp{{Solar Phys.}}   
\def\gafd{{Geophys. Astrophys. Fluid Dyn.}} 
\def\acta{{Acta Astron.}}
\def\jfm{{J. Fluid Mech.}}

\def\AIP{Astrophysikalisches Institut Potsdam}
\def\F{Ferri\`{e}re}
\def\R{R\"udiger}
\def\E{Elstner}
\def\K{Kitchatinov}
\def\B{Brandenburg}

\def\qq{\qquad\qquad}                      
\def\qqq{\qquad\qquad\qquad}               
\def\q{\qquad}
\def\DR{differential rotation\ }
\def\bib{\item{}}
\def\top{\item}
\def\toptop{\itemitem}
\def\so{$\Longrightarrow$\ }
\def\start{\begin{itemize}}
\def\stop{\end{itemize}}
\def\beg{\begin{equation}}
\def\ende{\end{equation}}
\def\ov{\bar}
\def\om{\omega}
\def\wt{\tilde}
\def\la{\langle}
\def\ra{\rangle}

\newcommand{\bs}{\\[12pt] }
\newcommand{\mfr}[2]{\frac{\displaystyle #1}{\displaystyle #2}}
\newcommand{\dv}[1]{\vec{\nabla} \cdot #1}
\newcommand{\ddt}[1]{\mfr{\partial #1}{\partial t}}
\newcommand{\gr}[1]{\vec{\nabla}#1}
\newcommand{\us}[1]{\mfr{1}{#1}}
\newcommand{\rat}[1]{\vec{\nabla} \times #1}
\newcommand{\str}[2]{\mfr{\partial #1}{\partial #2}}
\newcommand{\Ha}{\mbox{Ha}}
\newcommand{\Pm}{\mbox{Pm}}
\newcommand{\Co}{C_\Omega}
\newcommand{\ass}{\alpha_{\mbox{ss}}}
\newcommand{\rmin}{\tilde{r}_\Omega}
\newcommand{\rmax}{R}
\newcommand{\epm}{\varepsilon_{\mbox{\tiny Pm}}}

\title{MHD instability in differentially rotating cylindric flows}
\author{ G. R\"udiger \and Y. Zhang}
\offprints{G. R\"udiger}
\institute{\AIP,  An der Sternwarte 16, D-14482 Potsdam, Germany}

\date{\today}
\markboth{G. R\"udiger \& Y. Zhang: MHD shear flow turbulence}
{G. R\"udiger \& Y. Zhang: MHD shear flow turbulence}

\abstract{The possibility that the magnetic shear-flow instability (MRI, 
Balbus-Hawley instability) might give rise to turbulence in  
a cylindric Couette flow is investigated through numerical
simulations. The study is linear and the fluid flow is supposed
to be incompressible and differentially rotating with a simple
velocity profile with $\Omega = a+b/R^2$.
The simplicity of the model is counterbalanced by the fact 
that the study is fully global in all three spatial directions
with boundaries on each side; finite diffusivities are also
allowed. The investigation is also carried out for several 
values of the azimuthal wavenumber of the perturbations in order
to analyse whether non-axisymmetric modes might be preferred,
which might produce, in a non-linear extension of the study, 
a self-sustained magnetic field.\\
We find that with magnetic field the instability is easier to
be excited than without magnetic field. The critical Reynolds number for 
Pm$=1$ is of order 50 independent of  whether the nonmagnetic flow is stable 
or not. Also we find that i) the magnetic field strongly reduces the number of 
Taylor vortices, ii) the angular momentum is transported outwards and iii) for finite cylinders a netto 
dynamo-alpha effect  results which is negative for  the upper part of the cylinder and which is positive for the lower part of the cylinder.\\
For magnetic Prandtl number smaller than unity the critical Reynolds 
number scales with Pm$^{-0.65}$. If this was true even for very small 
magnetic Prandtl number (e.g. liquid sodium)  the critical Reynolds number 
should reach  the value 10$^5$ which is, however,  also characteristic for nonlinear finite-amplitude  hydrodynamic 
Taylor-Couette turbulence 
-- so that we easily have to expect the simultaneous existence of both sorts of instabilities in cold accretion disks.}

\maketitle

\begin{keywords}
Magnetohydrodynamis, accretion disks, turbulence 
\end{keywords}
\section{Introduction}
In recent years the  generation of turbulence in accretion disks is thought as basing on  the magnetorotational instability (MRI, 
Balbus-Hawley instability) in
which a differential rotation with outwards decreasing basic rotation rate becomes unstable under the influence of a given magnetic field (Velikhov 1959; Chandrasekhar 1961). After the
first work (Balbus \& Hawley 1991) several other papers  have studied the nonlinear evolution of the instability (Hawley \&
Balbus 1991; Brandenburg \ea  1995; Hawley \ea 1995; Matsumoto \&
Tajima 1995). By other  investigations  the importance of
boundaries  has been found (Curry \ea 1994; Curry \&
Pudritz 1995, 1996; Kitchatinov \& R\"udiger 1997; Kitchatinov \& Mazur
1997; R\"udiger \ea 1999). All the studies, however,  were linear, a fully global nonlinear approach
being just at the beginning (e.g. Drecker \ea 2000).

On the other hand, Dubrulle (1993), Richard \& Zahn (1999) and Duschl \ea (2000) focus attention 
to existence of instabilities in hydrodynamic Taylor-Couette flow, i.e. a 
configuration between two rotating cylinders (see also Koschmieder 1993). There 
we have to distinguish two cases. For sufficiently slow rotation of the outer 
cylinder and sufficiently rapid rotation of the inner cylinder there is indeed 
a linear instability, which, however, is not interesting for astrophysical 
applications such as for accretion disks. Here, however, one can ask 
for the nonlinear instabilities due to finite-amplitude disturbances. They indeed 
    exist, but for Reynolds numbers exceeding those for linear instability by 
    several orders of magnitudes. For a wide gap between both cylinders, the data discussed by 
Richard \& Zahn (1999) lead to a Reynolds number for finite-amplitude 
instability of about $2 \cdot 10^5$ (see below).

In the present paper we shall 
    demonstrate  for Taylor-Couette flow for plasma under the presence of 
    an external field that there are always linear instabilities for rather moderate 
    Reynolds numbers independent of whether the hydrodynamic Taylor-Couette flow 
    is unstable or not. We are able to compute the MHD instability for magnetic Prandtl numbers of
$0.01 \leq {\rm Pm} \leq 1$, however, for cold accretion disks and technical liquid metals like sodium the 
magnetic Prandtl numbers are much smaller.  For a  (realistic) Prandtl number of $10^{-5}$ an extrapolation is needed 
leading to a Reynolds number of about $10^{5}$. So we have the very interesting  situation that for 
natural electrically conducting fluids the critical Reynolds number for linear MHD-turbulence 
approaches the critical Reynolds number for  nonlinear hydrodynamic turbulence. If this is true then the 
theory of protoplanetary disks must be modified in an important sense. 

\begin{figure}
\hbox{
\psfig{figure=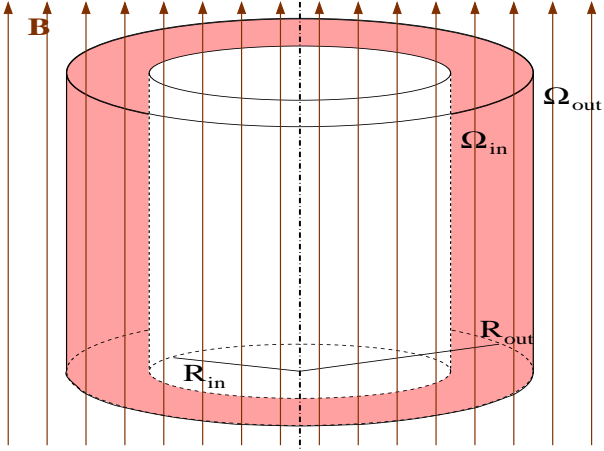,width=8cm,height=6cm}
}
\caption{Geometry of the cylindric Couette flow}
\label{geometry}
\end{figure}

We investigate here the effect of the instability in a simple numerical
model for a cylindric Taylor-Couette flow in which a magnetised fluid is 
contained between two finite cylinders. We consider the permissible
stationary circular flows of an incompressible fluid between two 
rotating coaxial cylinders (Fig. \ref{geometry}). As we shall  see, in the absence 
of viscosity, the class of such permissible flows is very wide: indeed,
if $\Omega$ denotes the angular velocity of rotation about the axis, then 
the equations of motion allow $\Omega$ to be an arbitrary function of the 
distance $R$ from the axis, provided the velocities in the radial and
the axial directions are zero. But if viscosity is present, the class
becomes  restricted: in fact, in the absence of any transverse 
pressure gradient, the most general form of $\Omega$ which is allowed is
\begin{equation}
\Omega(R) = a+{b\over {R}^2},
\label{angular}
\end{equation}
where $a$ and $b$ are two constants which are related to the angular
velocities $\Omega_{\rm in}$ and $\Omega_{\rm out}$ with which the inner and the outer
cylinders are rotated. Thus, if $R_{\rm in}$ and $R_{\rm out}$ ($>R_{\rm in}$) 
are the radii
of the two cylinders, then
\begin{equation}
\Omega_{\rm in}=a+b/R_{\rm in}^2, \;\;\;\;     
 \;\;\;\;      \Omega_{\rm out}=a+b/R_{\rm out}^2.
\label{inner}
\end{equation} 
Solving for $a$ and $b$ in terms of $\Omega_{\rm in}$ and $\Omega_{\rm out}$, we have
\begin{equation}
a=\Omega_{\rm in}{\hat\mu-\hat{\eta}^2 \over1-{\hat\eta}^2}, \;\;\;\;\;\;
 \;\;\;\;\;\;
b=\Omega_{\rm in}{{R_{\rm in}}^2(1-\hat\mu)\over1-{\hat\eta}^2},
\label{AB}
\end{equation}
where
$\hat\mu=\Omega_{\rm out}/\Omega_{\rm in}
$ and
$
\hat\eta=R_{\rm in}/R_{\rm out}$.

Flow and field perturbations are developed after the azimuthal
Fourier modes exp(i$m\phi$). Although the flow and field quantities are then complex numbers 
only their real parts have an own physical meaning.

\section{Equations and  model}
We solve the incompressible, dissipative, linearised MHD equations
 numerically in cylindrical coordinates for a global
model. The basic rotation law is given in (\ref{angular}) 
and the rotating fluid may  
be threaded by a strictly vertical magnetic field
\beg
 \vec{B_0} = B_0 \vec{e_z}
\label{1.1}
\ende 
(cf.
Stone \& Norman 1994). Extra radially dependent azimuthal
fields are also possible in this approximation (Curry \&
Pudritz 1995; Ogilvie \& Pringle 1996; Terquem \& Papaloizou 1996; Papaloizou \& Terquem
1997). By linearising around
the equilibrium state and using dimensionless quantities, our equations
read
\begin{eqnarray}
\ddt{\vec{u}} & = & \Co \left( - \vec{\cal{A}}
- \gr{p} \right) +  \left(\mfr{\Ha^2
\Pm}{\Co}\right)\vec{\cal{L}} + {\rm Pm}\ \Delta \vec{u} , 
\label{c} \bs
\Delta p & = & \dv{ \left( - \vec{\cal{A}} + \left(
\mfr{\Ha^2 \Pm}{\Co^2} 
\right) \vec{\cal{L}} + \mfr{\Pm}{\Co} \Delta \vec{u} \right) } ,
\label{n} \bs
\ddt{\vec{B}} & = & {\rm curl}{ \left(\Co \left(
\vec{\cal{R}} + \vec{\cal{E}} \right) - {\rm curl}{\vec{B}} \right) } ,  
\label{d}
\end{eqnarray}
where \( p \) is the pressure divided by the density, \( \vec{u} \) the velocity perturbation and
\( \vec{B} \) the magnetic perturbation. The normalisations are the same as in 
R\"udiger \ea (1999).

\begin{figure*}
\hbox{
\psfig{figure=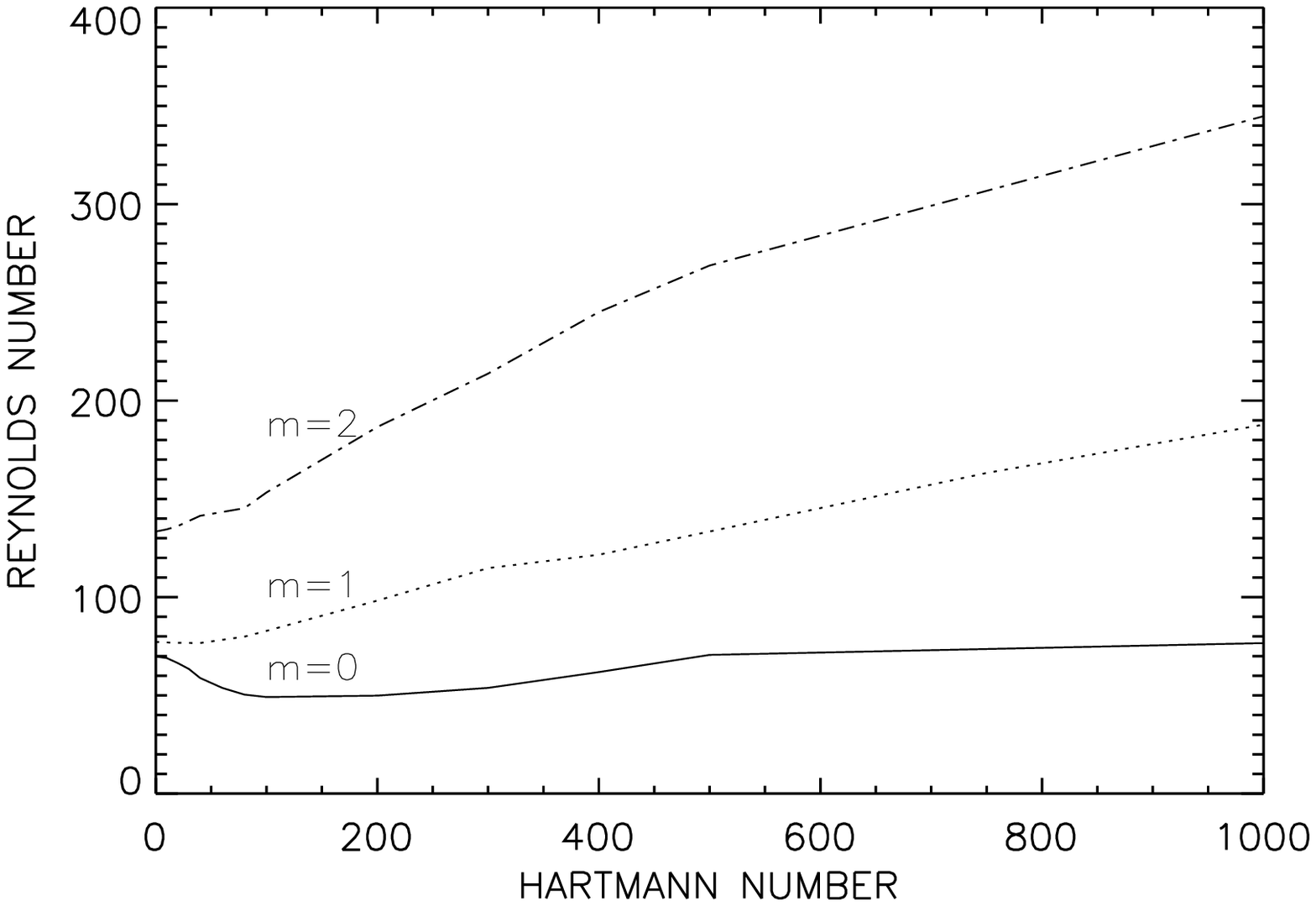,width=8cm,height=6cm}
\psfig{figure=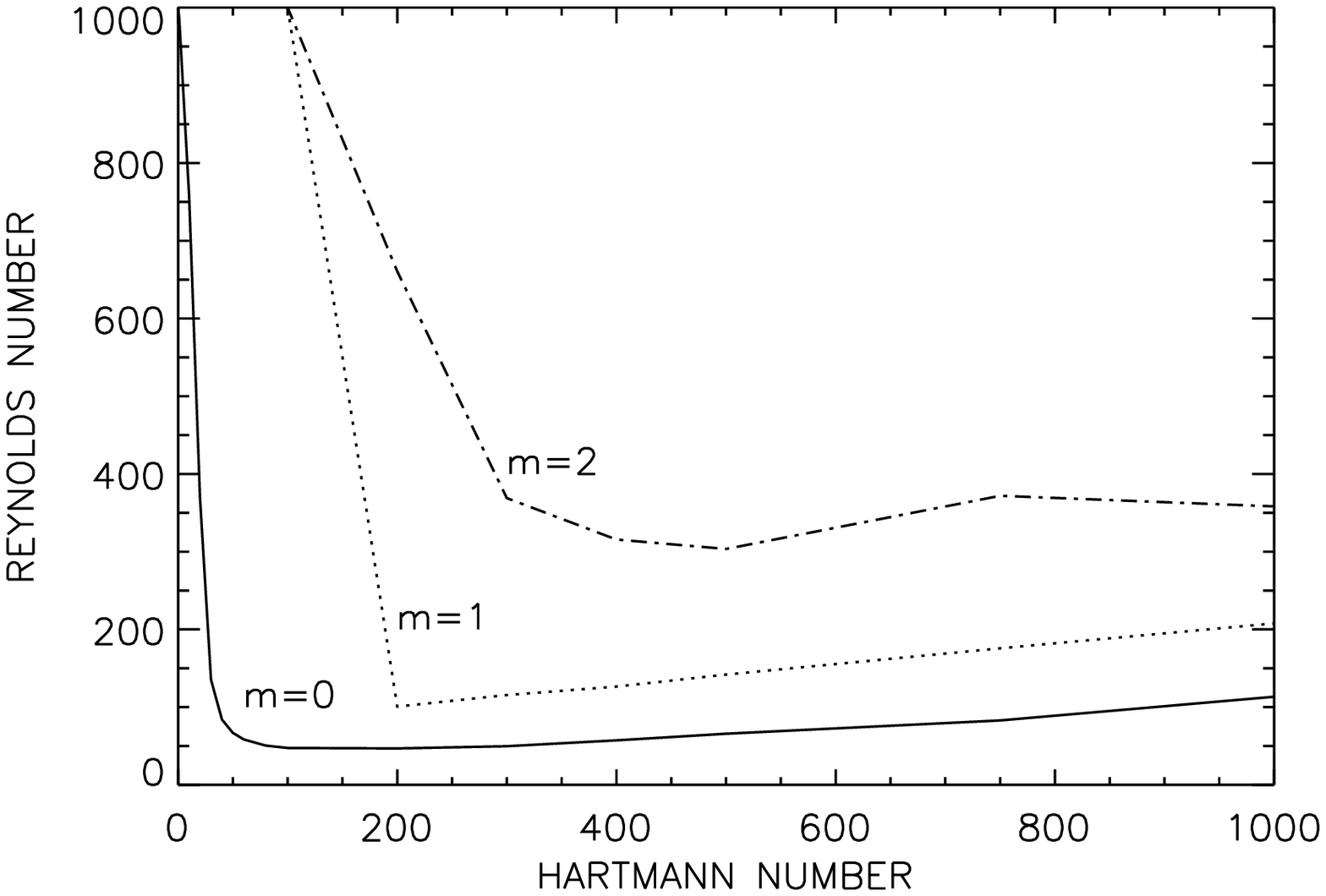,width=8cm,height=6cm}
}
\caption{Neutral-stability lines for various azimuthal wavenumbers ($m= 0,1,2$)
and rigid boundary conditions. The outer cylinder is at rest (left, 
hydrodynamic instability exists) or it rotates with 33\% of the rotation rate 
of the inner cylinder (right, hydrodynamic instability does not 
 exist). Pm$=1$, 
$\hat\eta = 0.5$}  
\label{ff1}
\end{figure*}

In the equations 
\( \vec{\cal{A}} \) and \( \vec{\cal{L}} \) are the linearised advection
term and Lorentz force given by
\begin{equation}
\vec{\cal{A}} = {\Omega\over \Omega_{\rm in}} 
\left(\matrix{{\rm i}mu_R - 2u_\phi \cr
{\rm i}mu_\phi + 2 {a\over \Omega} u_R \cr {\rm i}mu_Z \cr}\right)
\label{2}
\end{equation}
and
\begin{equation}
\vec{\cal{L}} = \left(\matrix{\str{B_R}{z}-\str{B_Z}{r} \cr
- \mfr{{\rm i}m}{r} B_Z + \str{B_\phi}{z} \cr
 0\cr} \right) , 
\label{3}
\end{equation}
while 
\beg
 \vec{\cal{R}} = r {\Omega\over \Omega_{\rm in}}
\left(\matrix{B_Z \cr 0 \cr -B_R\cr}\right)  
\label{3.1}
\ende
and 
\beg
 \vec{\cal{E}} = \left(\matrix{u_\phi \cr -u_R \cr 0\cr}\right)  
\label{3.2}
\ende
are the linearised electromotive force induced
by the basic differential rotation and the flow perturbations.
In the equations above small $r$'s  are the dimensionless
radius $r=R/H$. Finally,  
\begin{equation}
{\rm Ha} = {B_0 H \over \sqrt{\mu_0\rho\eta\nu}} 
\label{4}
\end{equation}
is the Hartmann 
number (\(\eta \), \( \rho \) and \( \nu \) being the magnetic diffusivity,
density and viscosity); 
\begin{equation}
\Co = {\Omega_{\rm in} H^2 \over \eta} 
\label{5}
\end{equation}
is the shear-flow parameter (`dynamo number') and \( \Pm \) is the 
magnetic Prandtl number, Pm$=\nu/\eta$. The quantity \( H \) is
the half-thickness of the cylinder which is also the unit length for the 
non-dimensionalization. The unit time is the magnetic diffusion time \(
\tau=H^2/\eta \), velocities are normalized with $H\Omega_{\rm in}$.
It is a time-stepping code using a global energy quenching
procedure of the form $C_\Omega =C_\Omega^{(0)}/(1+E)$
to find the lowest excitation regime, where $E$ is the total
energy of the system (see Elstner \ea 1990). 
In our computations we use another parameter named Reynolds number, it is
defined as
\begin{equation}
{\rm Re}={\Omega_{\rm in}R_{\rm in}(R_{\rm out}-R_{\rm in})\over\nu} ,
\label{Re}
\end{equation}
so finally we get
\begin{equation}
{\rm Re}={C_\Omega\hat\eta(1-\hat\eta)\over {\rm Pm}\Gamma^2}
\label{Re2}
\end{equation}
with $\Gamma = H/R_{\rm out}$. 

With our definitions the nonlinear instability 
described by Richard \& Zahn (1999) -- using experimental data of Wendt (1933) and Taylor (1936) -- 
 starts for
\beg
{\rm Re} \gsim 5\cdot 10^6 {\hat \eta (1-\hat\eta)^2 \over (1+\hat \eta)^3 
(1-\hat\mu)},
\label{RE}
\ende
a number which is about $2\cdot 10^5$ for $\hat \mu=0$ and $\hat \eta 
= 0.5$ used in our computations below.

A staggered mesh of 81 by 81 grid points
was used to solve the system.
In the Eqs. (\ref{c})--(\ref{d}) \( \Ha \), \( \Co \) and \( \Pm \)
are free parameters.
We solved the equations in order to get a representation of the
curves of marginal stability in the Ha-$C_\Omega$ plane. 

We  impose rigid boundary conditions for the velocity field,
both taken on {\em all} boundary layers and
set pseudo-vacuum boundary conditions on the magnetic field, \ie
the components of the magnetic flux tangential to the surface are vanishing. This 
choice also comes for simplicity in imposing the conditions for the 
flow, since by using a staggered mesh the grid-points for the velocity are
not defined directly on the boundaries. It also ensures the
necessary vanishing of the Poynting flux vector in normal
direction (cf. Ruzmaikin \ea 1988). It is generally accepted that
the MRI is influenced by the presence of the boundaries
(Curry \& Pudritz 1995).
It is thus of high importance to find
out how the various boundary conditions influence (say) the
symmetry of the solutions with respect to the rotation axis or
with respect to the equator. Local simulations for closed boxes
can not answer such questions with their high relevance to observations.

\section{Results}
\subsection{The neutral-stability lines}
The neutral-stability lines for $m< 3$ in the Re-Ha plane for ${\rm Pm}=1$
and various $\hat \mu$ are given in Fig. \ref{ff1}. 
The curves have an intersection with the $\rm Re$-axis;
for the line $m=0$, when ${\rm Ha}=0$, we can get ${\rm Re}=69.75$. 
This case is well-known for the
nonmagnetic flow, from Chandrasekhar (1961) we
find ${\rm Re}=68.3$. Both values are close to each other and 
our numerical simulations are consistent with the experiment.
There is also a minimum at ${\rm Re}=49.15$ for the line of $m=0$. 
We can conclude that 
with magnetic field the instability is easier to be excited than without
 magnetic field. For stronger magnetic field the critical Reynolds number again 
 increases, so that as known the instability is suppressed by high magnetic 
 field amplitudes.

In Fig. \ref{ff1} also the case of a flat rotation law is discussed where no 
hydrodynamical instability exists ($\hat \mu = 0.33$). There is now no eigenvalue 
along the vertical axis. But with magnetic field the minimum of the Reynolds 
number is hardly changed (${\rm Re}=46.9$) compared to the former case. It can be stressed that for 
Hartmann numbers of order 100 there is always a critical Reynolds number of order 
50 above which we have instability -- independent of whether or whether not a 
hydrodynamic instability exists. 

The results might be applied to two different 
cases. The first one is the solar tachocline with its values of 
    $\nu \simeq 10$ cm$^2$/s and $\eta \simeq 10^3$ cm$^2$/s (Pm$=10^{-2}$, see 
    R\"udiger \& 
    Kitchatinov 1996). The Reynolds numbers are much higher than $10^2$. It follows 
    $B_0 = 10^{-9} {\rm Ha}/\epsilon$ 
    Gauss with $\epsilon = H/R$ for the characteristic magnetic field strength. 
      With ${\rm Ha}\simeq 10^2$ and $\epsilon 
    \approx 0.1$ one finds $B_0 \simeq 1\ \mu$Gauss. 
    
  On the other hand, an experiment with fluid sodium with Pm$=10^{-5}$ ($\nu=10^{-2}$ cm$^2$/s, 
  $\eta= 10^3$ cm$^2$/s) 
  leads to $B_0\approx 10^{-1}$ Ha for $H\simeq 1$ m, i.e. $B_0 \simeq 10$ Gauss. 
  With $R_{\rm in}$ and $\Delta R$ of 10 cm each, it is Re$\simeq 10^5 f$ with the 
  rotation frequency $f$, so that it is no problem to exceed critical Reynolds 
  numbers of order 100.   

\subsection{The eigenfunctions}
\begin{figure}
\psfig{figure=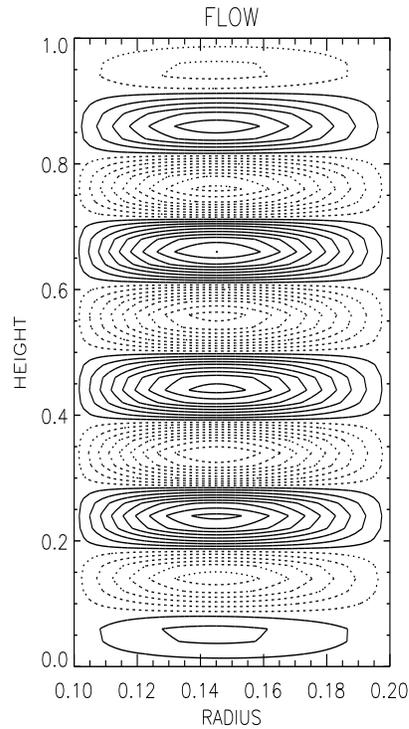,width=6cm,height=10cm}
\caption{The pattern  for the the azimuthal flow without magnetic field (Ha=0) but for resting outer cylinder ($\hat \mu=0$). 
$\hat \eta=0.5$, Pm=1, m = 0}
\label{ff8}
\end{figure}
\begin{figure*}
\hbox{
\psfig{figure=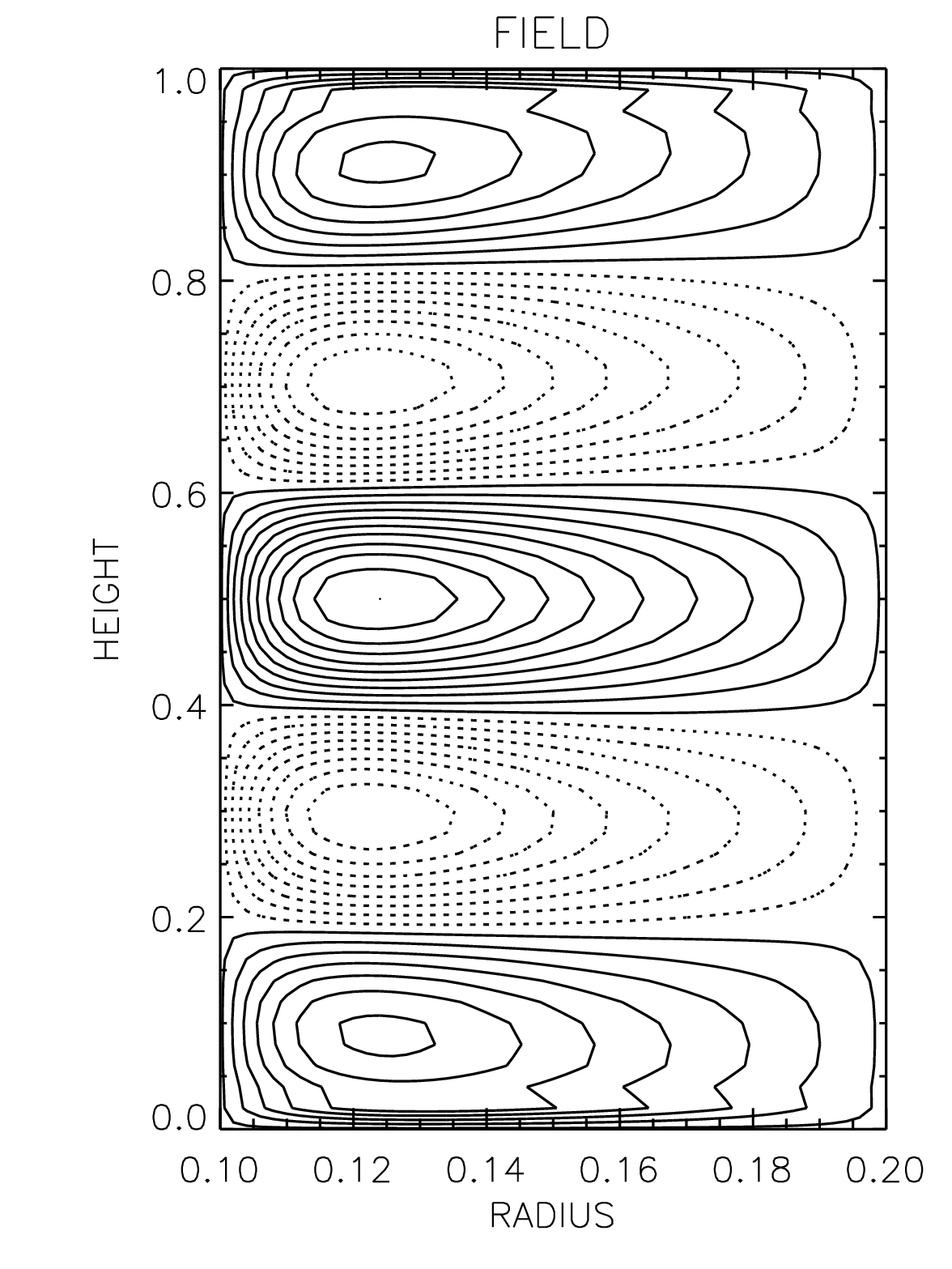,width=6cm,height=10cm}
\psfig{figure=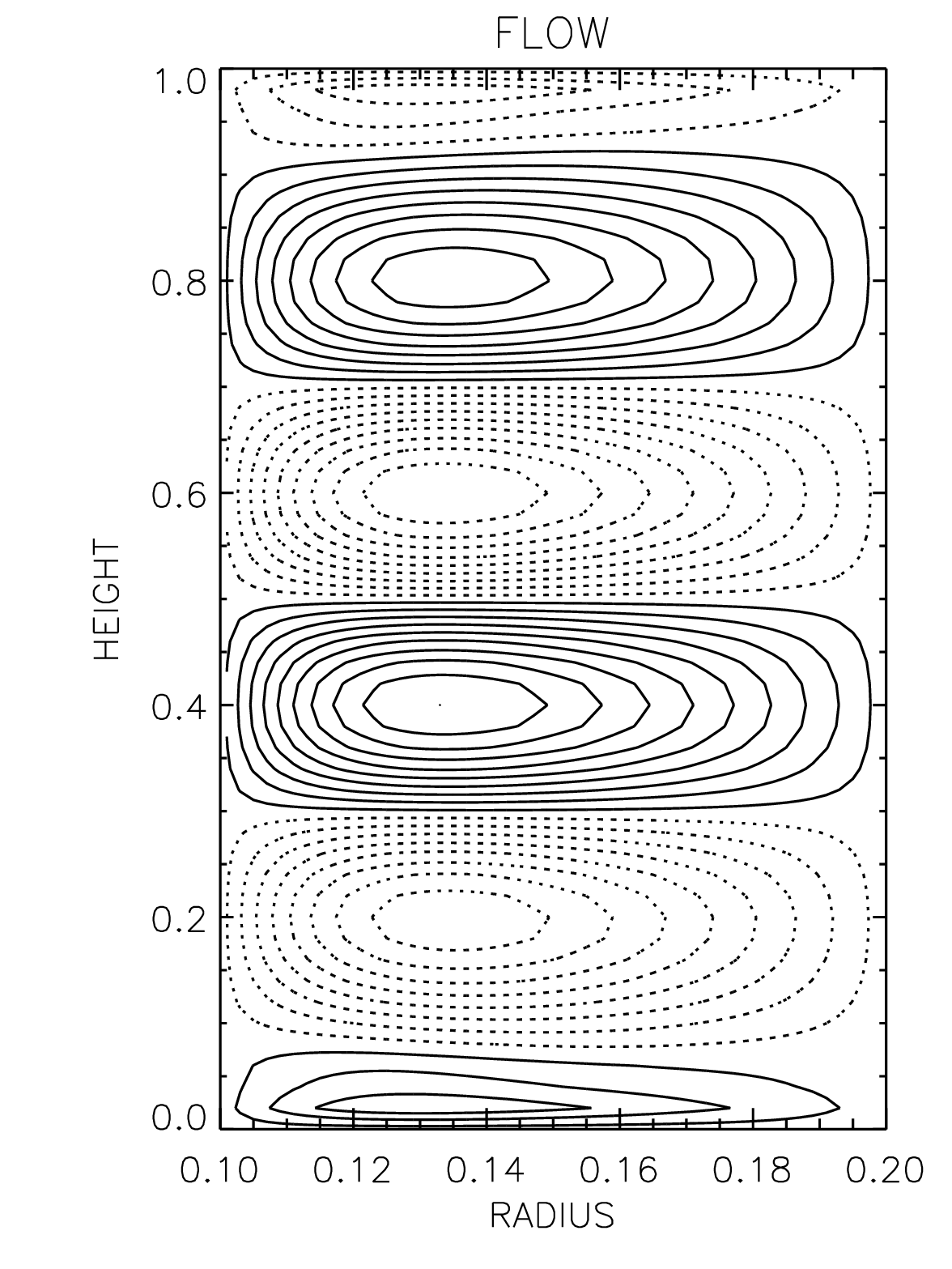,width=6cm,height=10cm}
}
\caption{The eigenfunction for the
toroidal magnetic field (left) and the azimuthal flow (right) in the gap between 
the cylinders. 
Ha $= 200$, Pm $= 0.1$, $m = 0$}
\label{ff9}
\end{figure*}

It is known  that the cell size in $z$ for the linear nonmagnetic 
Taylor-Couette flow
is practically the same as it is in $r$ -- so we can find 10 cells in 
Fig. \ref{ff8} for Ha$=0$. With magnetic field, 
however, only 5 cells develop antisymmetric with respect to the equator  
(Fig. \ref{ff9}). Obviously, the value of the 
vertical wave number strongly decreases for increasing  magnetic field (see Chandrasekhar 1961). 

In Fig. \ref{ff9} the induced toroidal magnetic field is also  shown.
It is an equatorially symmetric (`S') field.
Correspondingly, the associated zonal kinetic flow  has an
antisymmetric structure with respect to the equator. The induced fields and 
flows here are concentrated to
the inner boundary.

\subsection{Viscosity-alpha and dynamo-alpha}
Although our study is linear, it is worth analysing the efficiency  of
the instability for the angular momentum transport. The effective viscosity
that should bring the angular momentum transport is generally parameterised
through the quantity \( \ass \) (Shakura \& Sunyaev 1973), defined in terms
of the Reynolds and Maxwell stress tensors by
\[ \left\langle{ u_R u_\phi-\mfr{B_R B_\phi}{\mu_0 \rho}}\right\rangle =
- \nu_{\rm T} \ R {\partial \Omega \over \partial R}  \]
with the normalisation
\beg
\nu_{\rm T} = \alpha_{\rm SS} H^2 \Omega ,
\label{nu}
\ende
where $\nu_{\rm T}$ is called the eddy viscosity and $\alpha_{\rm SS}$ is  the viscosity-alpha. As we know, the $\alpha_{\rm SS}$ is not trivially positive, there are also hydrodynamic simulations with negative correlations $\left\langle{ u_R u_\phi}\right\rangle $.

In dimensionless units
\beg
\alpha_{\rm SS} \propto  \left\langle{ u_R u_\phi - {{\rm Ha}^2 \Pm
\over \Co^2} B_R B_\phi}\right\rangle .
\label{as}
\ende
In our linear approach the
amplitudes of the quantities are 
unknown as they are free of a common positive or negative
factor. Hence, the sign of quadratical terms can be computed as
well as ratios of fluctuating quantities.

The same concerns to the electromotive force $\vec{\cal{E}}= \langle \vec{u}' 
\times \vec{B}'\rangle$ which is due to the external field, ${\cal E}_z = 
\alpha_{zz}\  B_z$. Apart from positive factors we find
\beg
\alpha_{zz} \propto \langle u_R B_\phi - u_\phi B_R\rangle
\label{alfzz}
\ende
where the averaging procedure must distinguish between upper and lower part of the 
cylinder. As known the alpha-effect is antisymmetric with respect to the 
equator, the question is whether it is positive at the upper part and negative 
at the lower part or v.v.
\begin{figure}
\psfig{figure=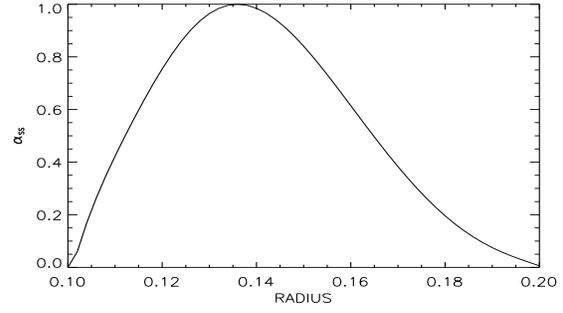,width=8cm,height=4.5cm}
\caption{The viscosity-alpha, $\alpha_{\rm SS}$, is always positive, i.e.  the  angular momentum 
transport is  outwards, even for the axisymmetric cells 
($m=0$). Ha$=200$, 
$\hat\mu = 0.33$, $\hat\eta = 0.5$ }
\label{ass}
\end{figure}

\begin{figure}
\psfig{figure=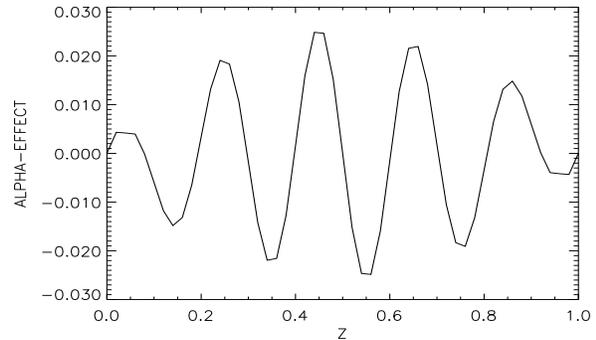,width=8cm,height=5cm}
\caption{The nonlinear quantity $\hat \alpha$ non-averaged as $z$-profile for 
$m=0$ and at a given radius. Pm$=0.1$, Ha$= 200$, $\hat\mu = 
0.33$, $\hat\eta= 0.5$ }
\label{alf}
\end{figure}

\begin{figure}
\psfig{figure=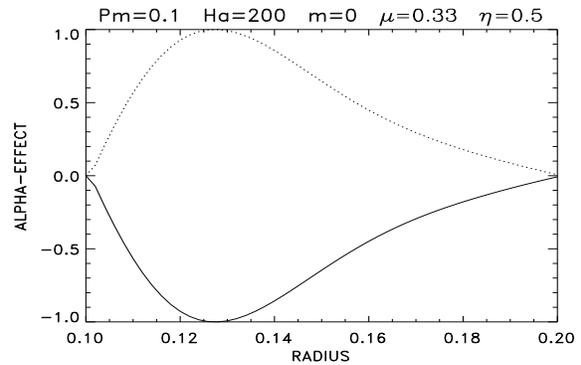,width=8cm,height=5cm}
\caption{The dynamo-alpha is negative in the upper hemisphere 
(solid) and positive in the lower hemisphere (dashed)}
\label{alpha}
\end{figure}

The results are presented in Figs. \ref{alf} and \ref{alpha}. Figure \ref{alf} for 
a prescribed radius displays the vertical profile of $\hat \alpha \equiv u_{\rm R} B_\phi 
- u_\phi B_{\rm R}$ for $m=0$ which has to be averaged to get the alpha-effect. The 
result is typical for the nature of this effect. The lower part of a cell 
has a negative (positive) $\hat \alpha$ in the upper (lower) part of the 
cylinder. The upper part of the cell is opposite. Due to the vertical stratification, however, the $\hat \alpha$ of the 
lower part of a cell dominates the $\hat \alpha$ of the upper part of a cell. 
The total alpha for each cell is negative (positive) for the upper (lower) part of 
the cylinder. The alpha-effect only exists if some kind of stratification exists. Figure \ref{alpha} 
gives the resulting alpha-effect after averaging $\hat \alpha$ about the entire 
upper part of the cylinder or the lower part, resp. Note that our computations lead 
to a negative (positive) total alpha-effect for the upper (lower) part of the 
cylinder. This result seems to be characteristic for the magneto-rotational 
instability (Brandenburg 1998) although we have been here no negative buoyancy. Obviously, for an infinite cylinder (or a model with 
periodic boundary conditions in $z$) there is no vertical stratification and, therefore, 
no alpha-effect.

\section{The magnetic Prandtl number dependence}
In order to approach real plasma we have to switch to magnetic Prandtl number 
smaller than unity. The solar gas beneath the convection zone approaches Prandtl 
numbers of order 10$^{-2}$ while liquid sodium due to its small viscosity with 
10$^{-5}$ is even more extreme. 

\begin{figure}
\hbox{
\psfig{figure=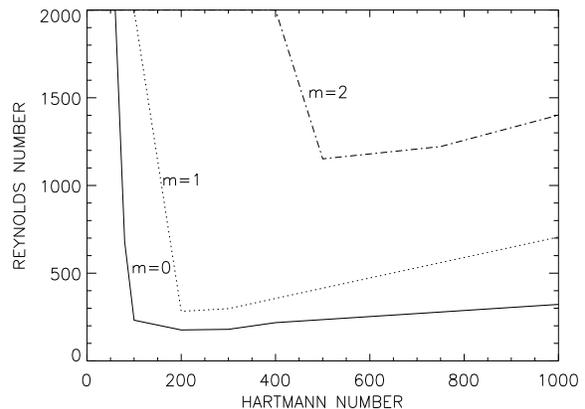,width=8cm,height=6cm}
}
\caption{Neutral-stability lines for various azimuthal wavenumbers 
($m = 0,1,2$) and rigid boundary conditions with Pm $= 0.1$. 
Hydrodynamic 
instability does not exist. The minimum Reynolds number is about 
 176. $\hat\mu = 
0.33$, $\hat\eta = 0.5$}
\label{ff01}
\end{figure}

\begin{figure}
\psfig{figure=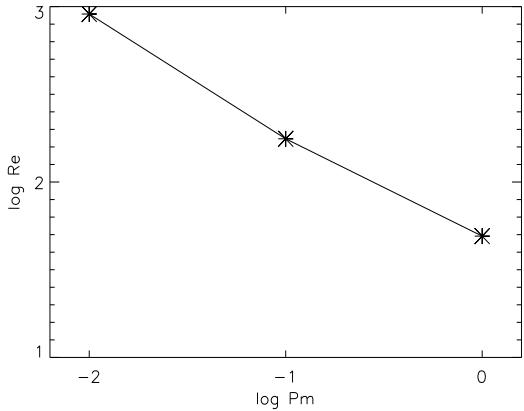,width=8cm,height=6cm}
\caption{Critical Reynolds numbers for various magnetic Prandtl numbers. The 
Reynolds number increases for decreasing Prandtl number. $m=0$}
\label{pm}
\end{figure}

Simulations with Pm$=0.1$ for a flat basic rotation law are given in detail in 
Fig. \ref{ff01}. The lowest eigenvalue is now about 171. Some computations with 
Pm$=0.01$ are also done. Figure \ref{pm} yields a slope such as 
\beg
{\rm Re}_{\rm crit} \propto {\rm Pm}^{-0.65},
\label{Recrit}
\ende
which we have only justified, however,  down to Pm$=0.01$. If it would also be true 
for  magnetic Prandtl numbers of order $10^{-5}$, the critical Reynolds number results as 9$\cdot 10^4$. 
    This value is very close to the Reynolds number which we know from 
    (\ref{RE}) for nonlinear but pure hydrodynamical instability. It might be the case that in the limit of very small magnetic Reynolds number, i.e. for very cold accretion disks we have the realisation of a {\em  mixed instability}: linear magneto-rotational instability and nonlinear Taylor-Couette flow turbulence.

\section{Discussion}
We have considered the Taylor-Couette flow experiment of a electrically 
conducting fluid under the presence of an external field.
It is  shown that for both flat or steep rotation law the MHD Taylor-Couette 
flow is always unstable. In contrast to the hydrodynamical regime the critical 
Reynolds number for instability does hardly depend on the rotation of the outer 
cylinder. The dependence of the critical Reynolds number on the magnetic Prandtl 
number, however, is much stronger, see Eq. (\ref{Recrit}). It is probably not too 
strong to avoid experimental work with liquid sodium. 

For Pm$=10^{-5}$ (liquid sodium) the estimated Reynolds number is of order $10^5$, 
which is also known from experiments concerning the nonlinear instability of 
Taylor-Couette flow. So we have the puzzling situation that the Reynolds number 
for two very different types of instability could be very similar, i.e. for 
    nonlinear hydrodynamic instability and linear magnetohydrodynamic instability. 
    If true, such a result will have strong implications for the simulations of 
    protoplanetary disks, which due to the low temperatures also have very low 
    magnetic Prandtl numbers.

Of particular interest for 
experiments also should be the existence of a turbulence electromotive force. Although its magnitude here 
remains open (as the  theory is linear) we can find the sign of the bilinear  expression. 
The EMF divided by the uniform  external field the dynamo-alpha  is found for the 
container. It is the $zz$-component of the alpha-tensor as only a vertical magnetic field is applied. 
The $\alpha_{zz}$ proves to be negative (positive) in the upper (lower) part of 
the cylinder. It only exists due to the finiteness of the container in vertical 
direction. One can understand this finding with the following. In order to 
construct the pseudo-tensor $\alpha$ one needs a scalar $\vec{g} \cdot \vec{\Omega}$ 
with a characteristic direction $\vec{g}$. In an vertically infinite cylinder the 
only characteristic direction $\vec{g}$ is the radial one, but $\vec{g} \cdot 
\vec{\Omega} = 0$ in this case. Only for cylinders which are bounded   in the vertical 
$\vec{g} \parallel \vec{\Omega}$ acts as an characteristic direction leading to 
a finite alpha-effect with an antisymmetry with respect to the equator.


\end{document}